\documentstyle[11pt]{article}
\title{
Phase separation and metal-insulator transitions in the spin-one-half
Falicov-Kimball model}
\author{Pavol Farka\v sovsk\'y\\
Institute  of  Experimental  Physics,  Slovak   Academy   of
Sciences\\
Watsonova 47, 043 53 Ko\v {s}ice, Slovakia}
\date{}
\begin{document}
\baselineskip=20pt
\maketitle

\begin{abstract}
The ground-state phase diagram of the spin-one-half Falicov-Kimball 
model (FKM) is studied in the one dimension using small-cluster 
exact-diagonalization calculations. The resultant exact solutions 
are used to examine possibilities for valence and metal-insulator 
transitions in this model. A number of remarkable results are found.
(i) The phase separation in the spin-one-half FKM takes place for a
wide range of $f$-electron concentrations $n_f$ and $d$-$f$ interactions
$G$, including $G$ large. (ii) In the strong coupling limit ($G>4$) the model
exhibits  a pressure induced discontinuous insulator-metal transition from 
an integer-valence state ($n_f=1$) into another integer-valence state 
($n_f=0$). (iii) For intermediate values of $G$ ($G\sim 2.5$) the FKM 
undergoes a few discontinuous intermediate-valence transitions. There are 
several discontinuous insulator-insulator transitions from $n_f=1$ to 
$n_f=1/2$ and a discontinuous insulator-metal transition from $n_f = 1/2$
to $n_f=0$. (iv) In the weak coupling limit ($G<2$) the model undergoes a few
consecutive discontinuous and continuous intermediate-valence transitions
as well as a discontinuous metal-insulator transition.

\end{abstract}

\newpage
\section{Introduction}
Recent new measurements of the insulator-metal transitions
in SmB$_6$~\cite{Cool} and in the transi\-tion-metal
halides~\cite{Pas},
have once again sparked an interest in valence and insulator-metal
transitions~\cite{Freericks}. These transitions are observed in a wide group 
of substances formed by transition-metal oxides as well as rare-earth
sulfides and borides, when some external parameters
(like pressure or temperature) are varied.
They are in many cases first-order phase transitions, however,
second-order transitions ranging from very gradual to rather
steep are also observed~\cite{Cho}.

To describe all such  transitions  in a unified picture
Falicov and Kimball~\cite{Fal} introduced a  simple  model in
which  only  two  relevant single-electron states  are  taken
into  account:  extended Bloch waves and a set of localized  states
centered  at  the sites of the metallic ions in the crystal.
It is assumed that insulator-metal transitions result from a change
in the occupation numbers of these electronic states, which remain
themselves basically unchanged in their character.
The Hamiltonian of the model can be written as the sum of four
terms:

\begin{eqnarray}
H_0=\sum_{ij\sigma}t_{ij}d^+_{i\sigma}d_{j\sigma}+
G\sum_{i\sigma\sigma'}f^+_{i\sigma}f_{i\sigma}
d^+_{i\sigma'}d_{i\sigma'}+ \nonumber \\
E_f\sum_{i\sigma}f^+_{i\sigma}f_{i\sigma}+
\frac{U}{2}\sum_{i\sigma}f^+_{i\sigma}f_{i\sigma}f^+_{i-\sigma}
f_{i-\sigma},
\end{eqnarray}
where $f^+_{i\sigma}$, $f_{i\sigma}$ are the creation and annihilation
operators  for an electron of spin $\sigma$ in the localized state at
lattice site $i$ with binding energy $E_f$ and $d^+_{i\sigma}$,
$d_{i\sigma}$ are the creation and annihilation operators of the
itinerant electrons in the $d$-band Wannier state at site $i$.

The first term of (1) is the kinetic energy corresponding to
quantum-mechanical hopping of the itinerant $d$ electrons
between sites $i$ and $j$. These intersite hopping
transitions are described by the matrix  elements $t_{ij}$,
which are $-t$ if $i$ and $j$ are the nearest neighbors and
zero otherwise (in the following all parameters are measured
in units of $t$).
The second term represents the on-site
Coulomb interaction between the $d$-band electrons with density
$n_d=N_d/L=\frac{1}{L}\sum_{i\sigma}d^+_{i\sigma}d_{i\sigma}$
and the localized $f$ electrons with density
$n_f=N_f/L=\frac{1}{L}\sum_{i\sigma}f^+_{i\sigma}f_{i\sigma}$,
where $L$ is the number of lattice sites. The third  term stands
for the localized $f$ electrons whose sharp energy level is $E_f$.
The last term represents the intra-atomic Coulomb interaction between
the localized $f$ electrons.

In spite of the fact that the FKM
is one of the  simplest examples of interacting fermionic system,
the theoretical picture of valence and insulator-metal transitions
remains still uncertain in the framework of this model.
Even, in the existing literature on this model, different answers
can be found on the fundamental question whether
the FKM can  describe both the discontinuous and continuous 
changes of the $f$ (d)-electron occupation number $n_f$ $(n_d)$
as a function of the $f$-level energy $E_f$~\cite{Sil}.
It should be noted that this question is indeed crucial for the systems
mentioned above, since, supposing~\cite{Cho} that the external pressure
shifts the energy level $E_f$, the valence changes observed in some
rare-earth and transition-metal compounds (SmS, SmB$_6$,
Ti$_2$O$_3$, and so on) could be understandable purely electronically.
Unfortunately, it was found that valence and insulator-metal
transitions are very sensitive to the approximation used.
Various approximations~\cite{Sil} (mean-field, virtual crystal,
CPA, etc.) yield very different and often fully controversial
results. This indicates that the study of valence and
insulator-metal transitions may be successful only with methods
which are relatively insensitive  to  the type of approximation
used and, of course, with  exact  methods.

In our previous papers~\cite{Far1,Far2} we have showed that
the method of small-cluster exact-diagonalization calculations
is very effective in describing ground-state properties
of both the spinless and spin-one-half FKM. For the spinless FKM~\cite{Far1} 
the exact numerical calculations (over the full set of $f$-electron
configurations) can be performed on relatively large clusters
($L\sim 36$), even without some special computational effort.
For such clusters the finite-size effects are practically negligible
and the results can be satisfactory extrapolated to the thermodynamics
limit ($L\to \infty$). Using this method we have described successfully
the ground-state phase diagram as well as the picture of valence and
metal-insulator transitions in the spinless FKM~\cite{Far1}. 
The situation for the spin-one-half FKM is more complicated. 
The full Hilbert space of the spin-one-half FKM is much larger than 
one of the spinless FKM and this fact impose severe restrictions on 
the size of clusters that can be studied by the exact-diagonalization 
method. Our recent~\cite{Far2} numerical
computations performed for the spin-one-half FKM with $U$ and $G$ finite
showed that clusters with $L>24$ are beyond the reach of present day
computers. Fortunately, the size of the Hilbert space can be reduced
considerably in some special, but physically still interesting limits,
e.g., $U\to \infty$. In the limit $U\to \infty$ states with two
$f$ electrons at the same site are projected out thereby much larger
clusters $(L\sim 36)$ become accessible for the numerical investigation
in this reduced subspace. For this reason all calculations presented
in this paper have been done at $U=\infty$.
The main goal for performing these calculations was to construct
the comprehensive phase diagram of the spin-one-half FKM. The second goal
of our numerical study was to find and describe all possible types of 
 valence and metal-insulator transitions in this model. We show that both
the ground-state phase diagram as well as the picture of valence and
metal-insulator transitions obtained for $U = \infty$ strongly differ
from ones obtained in our previous paper~\cite{Far2} for a restricted set 
of $f$-electron configurations and $U$ finite.
In particular, the exhaustive numerical studies of the model
(at $U=\infty$) performed on finite clusters up to $36$ sites revealed
some new unexpected features like the phase separation for all nonzero $G$ 
and discontinuous metal-insulator transitions for $G$ small, 
to mention only a few. In the present paper we discuss these features in 
detail.

Since the $f$-electron density operators
$f^+_{i\sigma}f_{i\sigma}$ of each site $i$ commute with
the Hamiltonian (1), the $f$-electron occupation number 
is a good quantum number, taking only two values, $w_i=0,1$
according to whether the site $i$ is unoccupied or occupied
by the localized $f$ electron (configurations with $w_i=2$
are projected out due to $U= \infty$).
Therefore the Hamiltonian (1) can be replaced by

\begin{equation}
H=\sum_{ij\sigma}h_{ij}d^+_{i\sigma}d_{j\sigma}+
E_f\sum_{i}w_i,
\end{equation}
where $h_{i,j}=t_{ij}+Gw_i\delta_{ij}$.

Thus for a given $f$-electron configuration
$w=\{w_1,w_2 \dots w_L\}$ defined on a one-di\-men\-sional
lattice with periodic boundary conditions, the Hamiltonian (2)
is the second-quantized version of the single-particle
Hamiltonian $h(w)=T+GW$, so the investigation of
the model (2) is reduced to the investigation of the
spectrum of $h$ for different configurations of $f$ electrons.
Since the $d$ electrons do not interact among themselves, the
numerical calculations precede directly in the following steps
(we consider only the case $N_f+N_d=L$,
which is the point of the special interest for valence
and metal-insulator transitions caused by promotion of electrons
from localized $f$ orbitals $(f^n \to f^{n-1})$ to the conduction
band states).
(i) Having $G$, $E_f$ and $w=\{w_1,w_2 \dots w_L\}$ fixed, find
all eigenvalues $\lambda_k$ of $h(w)=T+GW$. (ii) For a given
$N_f=\sum_iw_i$ determine the ground-state energy
$E(w)=\sum_{k=1}^{L-N_f}\lambda_k+E_fN_f$ of a particular
$f$-electron configuration $w$ by filling in the lowest
$N_d=L-N_f$ one-electron levels (the spin degeneracy must be taken
into account). (iii) Find the $w^0$ for which
$E(w)$ has a minimum. Repeating this procedure for different
values of $N_f,G$ and $E_f$, one can immediately study
the ground-state phase-diagram of the model in the $N_f$-$G$ plane
as well as the dependence of the $f$-electron occupation number
$N_f=\sum_iw^0_i$ on the $f$-level position $E_f$ (valence transitions).

\section{Phase separation}
Let us first discuss the ground-state phase-diagram of the spin-one-half
FKM. To reveal the basic structure of the phase diagram in the $N_f$-$G$
plane ($E_f=0$) we have performed an exhaustive study of the model on finite
(even) clusters up to 36 sites. For fixed $L$ the numerical calculations
have been  done along the lines discussed above with a step $\Delta N_f=2$
and $\Delta G =0.05$. The results of numerical computations
are summarized in Fig.~1. These results show that the phase diagram of the 
spin-one-half FKM consists of three main domains: 
the most homogeneous domain MHD~(in Fig.~1 denoted as $\cdot$) and two phase 
separation domains $PSD_1$~(denoted as $\circ, \times$ and $+$) 
and $PSD_2$~(denoted as $\triangle$).
In the MHD the ground states are configurations in which the atomic or 
$n$-molecule clusters of $f$-electrons are distributed in such a manner that 
the distances between two consecutive clusters are either $d$ or
$d+2$. Furthermore the distribution of the distances of $d$ and $d+2$ has 
to be most homogeneous. Two basic types of the ground state configurations 
that fill up practically the whole MHD are displayed in Table.~1. 
In the $PSD_1$ the ground state configurations are mixtures $w\&w_e$
of aperiodic (periodic) configurations $w$ and the empty configuration
$w_e=\{00 \dots 0\}$~(see Ref.~\cite{Gruber} for a definition of mixtures), 
i.e., all $f$-electrons are distributed only in one part of the lattice 
(configuration $w$) while another (connecting) part is free of $f$-electrons
(phase separation). We have found three basic types of configurations $w$ 
which form these  mixtures: (i) the aperiodic atomic configurations
$w_a=\{10_{k_1}10_{k_2}\dots10_{k_i}1\dots0_{k_2}10_{k_1}1\}$
(with $k_i>0$), (ii) the aperiodic $n_i$-molecule configurations
$w_b=\{1_{n_1}0_{k_1}1_{n_2}0_{k_2}\dots1_{n_i}0_{k_i}1_{n_i}\dots
0_{k_2}1_{n_2}0_{k_1}1_{n_1}\}$ (with $1\leq n_i\leq N_f/2$ and $k_i>0$),
and (iii) the $N_f$-molecule (segregated) configurations $w_c=\{11\dots 1\}$.
Three different regions of stability corresponding to mixtures
$w_a\&w_e$, $w_b\&w_e$ and $w_c\&w_e$ are denoted in Fig.~1 as 
$\circ, \times$ and $+$.
It is seen that the mixtures of the atomic configuration $w_a$
and the empty configuration are stable only at low $f$-electron
concentrations and the Coulomb interactions $G<2.2$.
A direct comparison of results obtained for the spin-one-half and spinless
FKM~\cite{Gruber} shows that this region corresponds roughly to a region of 
phase separation in the spinless FKM. Outside these regions the phase diagrams
of the spin-one-half and spinless FKM are, however, strongly different.
While the phase separation in the spinless FKM takes place only for
weak interactions ($G<1.2$) the spin-one-half FKM exhibits the phase
separation for all Coulomb interactions. Even with increasing $G$
the phase separation shifts to higher $f$-electron concentrations.
Particularly, in the region $2.5<G<2.7$ where the ground states are
the mixtures of the $n$-molecule configurations $w_b$
with the empty configuration the phase separation takes place for
all $N_f<L/2$ and in the region $G>2.7$ where the ground states are
the segregated configurations $w_S=w_c\&w_e$ even for all $N_f<L$. 
At large $f$-electron concentrations $n_f$
but in the opposite limit $(G<0.4)$ there exists another small domain
of the phase separation $PSD_2$ (denoted as $\triangle$). 
The numerical results on finite lattices up to 36 sites 
revealed only one type of configurations that can be the ground
state configurations in this domain, and namely, the mixtures of the periodic
$n$-molecule configurations $w_d$ with the fully occupied lattice 
$w_f=\{11\dots 1\}$ (the length of a connected cluster of occupied sites 
in these mixtures is at least $L/2$). 

The second step in our numerical studies has been the extrapolation
of small-cluster exact diagonalization results on large lattices.
In Fig.~2 we present the ground state phase diagram of the spin-one-half
FKM obtained for $L=100$ on the extrapolated set of configurations that
includes practically all possible types of the ground-state configurations
found on finite lattices up to 36 sites. In particular,
we have considered (i) the most homogeneous configurations $w_h$ of the 
type $a$ and $b$ (see Table~1), (ii) all mixtures $w_a\&w_e$ with $k_i$ 
smaller than 6, (iii) all mixtures $w_b\&w_e$ with $n_i$ and $k_i$ smaller 
than 6, (iv) all mixtures $w_d\&w_f$ with periods smaller than 12, 
and (v) all segregated configurations.
One can see that all fundamental features of the phase diagram found
on small lattices hold on much larger lattices too. 
Of course, the phase boundaries  of different regions corresponding 
to mixtures $w_a\&w_e$ $(a)$, $w_b\&w_e$ $(b)$, $w_c\&w_e$ $(c)$ and 
$w_d\&w_f$ $(d)$ are now more obvious.  
Since the mixtures $w_a\&w_e$, $w_b\&w_e$, $w_c\&w_e$ and $w_d\&w_f$ are
metallic~\cite{Gruber} one can expect that the phase boundary between 
the MHD and PSD is also the boundary  of the correlation induced 
metal-insulator transition. To confirm this conjecture it is necessary
to show that the most homogeneous configurations from the MHD are insulating,
or by other words that there is a finite energy gap $\Delta$ at the Fermi
energy in the spectra of these configurations~\cite{note}. The numerical 
results for the most-homogeneous configurations of the type $a$ and $b$ that 
fill up practically the whole MHD are displayed in Fig.~3. The insulating
character of these configurations is clearly demonstrated by the finite
$\Delta$ that exists for all nonzero $f$-electron concentrations $n_f$ 
and Coulomb interactions $G$. Thus we can conclude that
the spin-one-half FKM undergoes (on the phase boundary between the
MHD and PSD) the correlation induced metal-insulator transition that
is accompanied by a discontinuous change of the energy gap $\Delta$.

\section{Valence and metal-insulator transitions}
The existence of a large metallic domain in the strong coupling region is the
main difference between the phase diagram of the spin-one-half and spinless
FKM. In the spinless FKM~\cite{Gruber} the existence of the metallic phase was
restricted only on a small region $G<1.2$ and $n_f<1/4$ ($n_f>3/4)$,
while the remaining part of the phase diagram was insulating.
The phase diagram of the spin-one-half FKM has a more complicated structure.
In addition to the insulating phase corresponding to the
homogeneous configurations $w_h$ there are four metallic phases
corresponding to four different classes of the ground state configurations:
$w_a\&w_e$, $w_b\&w_e$, $w_c\&w_e$ and $w_d\&w_f$. Of course, this fact has 
to lead also to different picture of valence and metal-insulator transitions 
induced by pressure (increasing $E_f$). Let us now discuss, in more
detail, possible types of these transitions. In order to examine possible types
of valence and insulator-metal transitions induced by pressure we have chose
three different values of $G$ $(G=1,2.5,5)$ that represent three typical
regimes of the spin-one-half phase diagram.
For $G=5$ the valence transition can be constructed immediately.
Indeed the ground states in this region are only the segregated configurations
and thus they can be used directly in numerical calculations instead of the
full set of $f$-electron configurations. This allows to perform numerical
calculations on large systems  and practically fully exclude finite size
effects. Results of numerical calculations are shown in Fig.~4.
It is seen that in the strong coupling limit the spin-one-half FKM  exhibits
a discontinuous valence transition from an integer valence ground state
$n_f=1$ into another integer valence ground state $n_f=0$ at $E_f=E_c\sim
1.273$. The inset in Fig.~4 shows that the discontinuous valence transition
is accompanied by a discontinuous insulator-metal transition since
the energy  gap $\Delta$ vanishes discontinuously at $E_f=E_c$.
Thus we can conclude that the spin-one-half FKM in the pressure
induced case can describe the discontinuous insulator-metal
transitions from an integer-valence state ($n_f=1$) into another
integer-valence state ($n_f=0$).

The situation for $G=2.5$ is slightly complicated. Although we know that the
ground states for $2.45<G<2.75$ (and $n_f<1/2$) are configurations of a type
$w_b\&w_e$ the number of configurations belonging to this class is still too
large for numerical calculations on large lattices and thus it should be
further reduced.  For this reason we have performed numerical calculations
for all finite (even) clusters up to 48 sites at selected value $G=2.5$.
We have found that for any $N_f>0$ only one type of configurations and namely
mixtures $\{11001100 \dots 1100\}\&w_e$ are the ground states at $G=2.5$.
This allows us to avoid  technical difficulties associated
with a large number of configurations and consequently to study very
large systems ($L\sim 1200$). The valence transition obtained for this set
of configurations (and $w_h$ configurations for $n_f>1/2)$
is shown in Fig.~5. It is seen that the valence transition for $G=2.5$
is very steep. A more detail analysis showed, however, that the transition
from $n_f=1$ to $n_f=0$ is not a discontinuous integer valence transition
but it consists  of several discontinuous intermediate valence transitions.
Particularly, there are several discontinuous insulator-insulator transitions
from $n_f=1$ to $n_f=1/2$ and a discontinuous insulator-metal transition
from $n_f=1/2$ to $n_f=0$. To exclude the possibility that this structure
is a consequence of the finite size of clusters used in numerical calculations
we have performed calculations on several large clusters ($L=400, 800, 1200$)
but no finite size effects on the valence transition were observed.

The most complicated situation is for small and intermediate values of $G$.
For example, at selected value of $G=1$ the ground states are the insulating 
configurations $w_h$ (for $n_f>0.16$) and the metallic configurations 
$w_a\&w_e$ (for $n_f<0.16$). However, for a construction of the valence 
transition on large lattices this knowledge is too general and thus 
for $n_f<n_c$ it is again necessary to perform an additional study of the 
model in order to determine the explicit type of ground-state configurations 
at selected value of $G$.  The exhaustive numerical studies that we have 
performed for $n_f<n_c$ on finite lattices up to 60 sites shoved that only 
configurations of the type 
$w_1=\{1010_{n_1}1010_{n_2}\dots 0_{n_2}1010_{n_1}101\}\&w_e$
or configurations of the type
$w_2=\{10_{n_1}1010_{n_2}1010_{n_3}\dots 0_{n_3}1010_{n_2}1010_{n_1}1\}\&w_e$
can be the ground states at $G=1$. This fact allows us to use the set
of configurations $w_1$ and $w_2$ instead of the much larger set $w_a\&w_e$
and consequently to construct the valence transition on
large lattices ($L\sim 1200$). The results of numerical calculations
obtained for the $E_f$-dependence of the $f$-electron occupation number
$n_f$ and the energy gap $\Delta$ are displayed in Fig.~6.  Unlike the
valence transitions obtained for $G=5$ and $G=2.5$
the valence transitions in the weak coupling limit are much wider
and consist of several consecutive discontinuous and continuous valence
transitions. The valence transitions for $E_f<E_c\sim -1.05$
are insulator-insulator transitions since they realize between
insulating ground states corresponding to the most homogeneous
configurations $w_h$. At $E_f=E_c$ the spin-one-half FKM undergoes
a pressure induced discontinuous valence transition from an intermediate
valence state with $n_f\sim 0.19$ into another intermediate-valence state
with $n_f\sim 0.14$ that is accompanied by a discontinuous insulator-metal
transition. Above $E_c$ the $f$-electron occupation number $n_f$ changes
continuously and vanishes at $E_f=E_0\sim -0.8$.

It should be noted that the weak-coupling picture of valence and
metal-insulator transitions presented in this paper strongly differs
from one found in our previous work~\cite{Far2} for the restricted set of 
configurations consisting of only the most homogeneous configurations $w_h$. 
For such a set of configurations the system exhibits the metal-insulator 
transition only at $E_f=E_0$ and this transition is continuous. This fact 
indicated the serious deficiency of the spin-one-half FKM since in some 
rare-earth compounds (e.g. in SmB$_6$~\cite{Cool}) just a discontinuous 
insulator-metal transition is observed. 
Fortunately, a more accurate analysis performed in the present
paper has not confirmed this deficiency of the model. Contrary, our numerical
results showed that the spin-one-half FKM is capable of describing not only
a discontinuous metal-insulator transition but all basic types of
valence transitions observed experimentally in rare-earth and
transitions-metal compounds. This indicates that the spin-one-half FKM
could yield the correct physics for describing these materials.

In summary, the extrapolation of small-cluster exact-diagonalization 
calculations has been used to examine the ground-state phase diagram of the 
spin-one-half FKM in the one dimension. A number of remarkable results have 
been found. (i) The phase separation in the spin-one-half FKM takes place for 
a wide range of $f$-electron concentrations $n_f$ and $d$-$f$ interactions
$G$, including $G$ large. (ii) In the strong coupling limit ($G>4$) the model
exhibits  a pressure induced discontinuous insulator-metal transition from 
an integer-valence state ($n_f=1$) into another integer-valence state 
($n_f=0$). (iii) For intermediate values of $G$ ($G\sim 2.5$) the FKM 
undergoes a few discontinuous intermediate-valence transitions. There are 
several discontinuous insulator-insulator transitions from $n_f=1$ to 
$n_f=1/2$ and a discontinuous insulator-metal transition from $n_f = 1/2$
to $n_f=0$. (iv) In the weak coupling limit ($G<2$) the model undergoes a few
consecutive discontinuous and continuous intermediate-valence transitions
as well as a discontinuous metal-insulator transition.

\vspace{0.5cm}
This work was supported by the Slovak Grant Agency VEGA
under grant No. 2/4177/97.

\newpage
Figure Captions

\vspace{0.5cm}
Fig. $1.$ 
The ground-state phase diagram of the spin-one-half FKM
obtained over the full set of $f$-electron configurations.
For $1/4<n_f<3/4$ numerical calculations have been done 
on the lattice with $L=28$ while for $n_f\leq 1/4$ and 
$n_f\geq 3/4$ on the lattice with $L=36$.
Four different regions of stability corresponding to mixtures
$w_a\&w_e$, $w_b\&w_e$, $w_c\&w_e$ and $w_d\&w_f$  are denoted as 
$\circ, \times,+$ and $\triangle$.

\vspace{0.5cm}
Fig. $2.$ 
The ground-state phase diagram of the spin-one-half FKM
obtained on the extrapolated set of $f$-electron configurations
for $L=100$.
Four different regions of stability corresponding to mixtures
$w_a\&w_e$, $w_b\&w_e$, $w_c\&w_e$ and $w_d\&w_f$  are denoted as 
$a,b,c$ and $d$. The dashed line through the exact numerical points is 
a guide to the eye.

\vspace{0.5cm}
Fig. $3.$ $n_f$-dependence of the energy gap $\Delta$ 
calculated for the most homogeneous configurations of the 
type $a$ and $b$ (see Table~1).

\vspace{0.5cm}
Fig. $4.$ Dependence of the $f$-electron occupation number $n_f$
on the $f$-level position $E_f$ for $G=5$ and three different 
values of $L$. Inset: The behavior of $n_f$ and $\Delta$ close
to the insulator-metal transition point.

\vspace{0.5cm}
Fig. $5.$ Dependence of the $f$-electron occupation number $n_f$
on the $f$-level position $E_f$ for $G=2.5$ and three different 
values of $L$. Inset: The behavior of $n_f$ and $\Delta$ close
to the insulator-metal transition point.

\vspace{0.5cm}
Fig. $6.$ Dependence of the $f$-electron occupation number $n_f$
on the $f$-level position $E_f$ for $G=1$ and three different 
values of $L$. Inset: The behavior of $n_f$ and $\Delta$ close
to the insulator-metal transition point.

\newpage
Table Captions
\begin{table}[h]
\caption{Two basic types of the most homogeneous configurations 
that fill up practically the whole MHD for  $L=24$.} 
\begin{center}
\begin{tabular}{|lll|}
\hline
$N_f$            & $w^a_h$& $w^b_h$\\
\hline
4  & 100001000000100001000000 &110000000000110000000000\\
6  & 100100001001000010010000 &110000001100000011000000\\
8  & 100100100100100100100100 &110000110000110000110000\\
10 & 110010011001001001100100 &110011000011001100110000\\
12 & 110011001100110011001100 &110011001100110011001100\\
14 & 111001110011100111001100 & \\
16 & 111100111100111100111100 & \\
18 & 111111001111110011111100 & \\
20 & 111111111100111111111100 & \\
\hline
\end{tabular}
\end{center}
\end{table}


\begin{thebibliography}{999}
\bibitem{note} Since we consider the case $N_f+N_d=L$, 
the Fermi level $E_F$ and the energy gap $\Delta$
of some configuration $w$ are given by $E_F=\lambda_{L-N_f}$ and
$\Delta=\lambda_{L-N_f+1}-\lambda_{L-N_f}$, respectively.
\end{thebibliography}
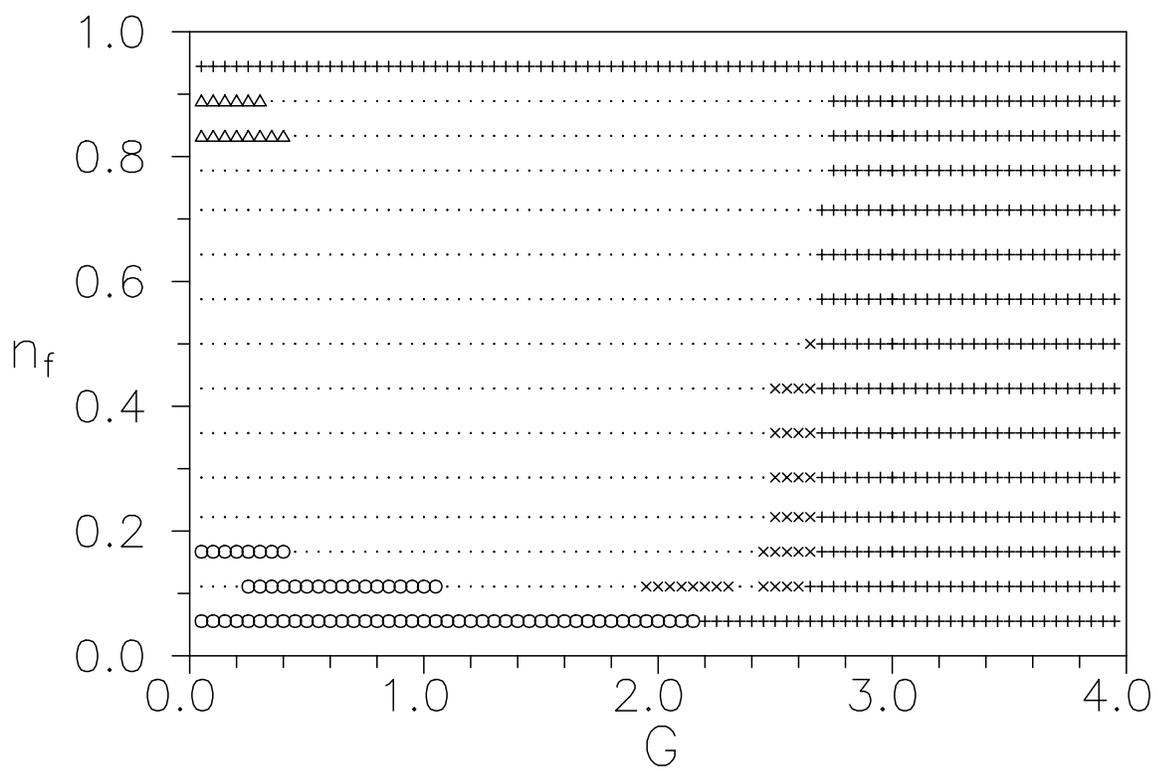
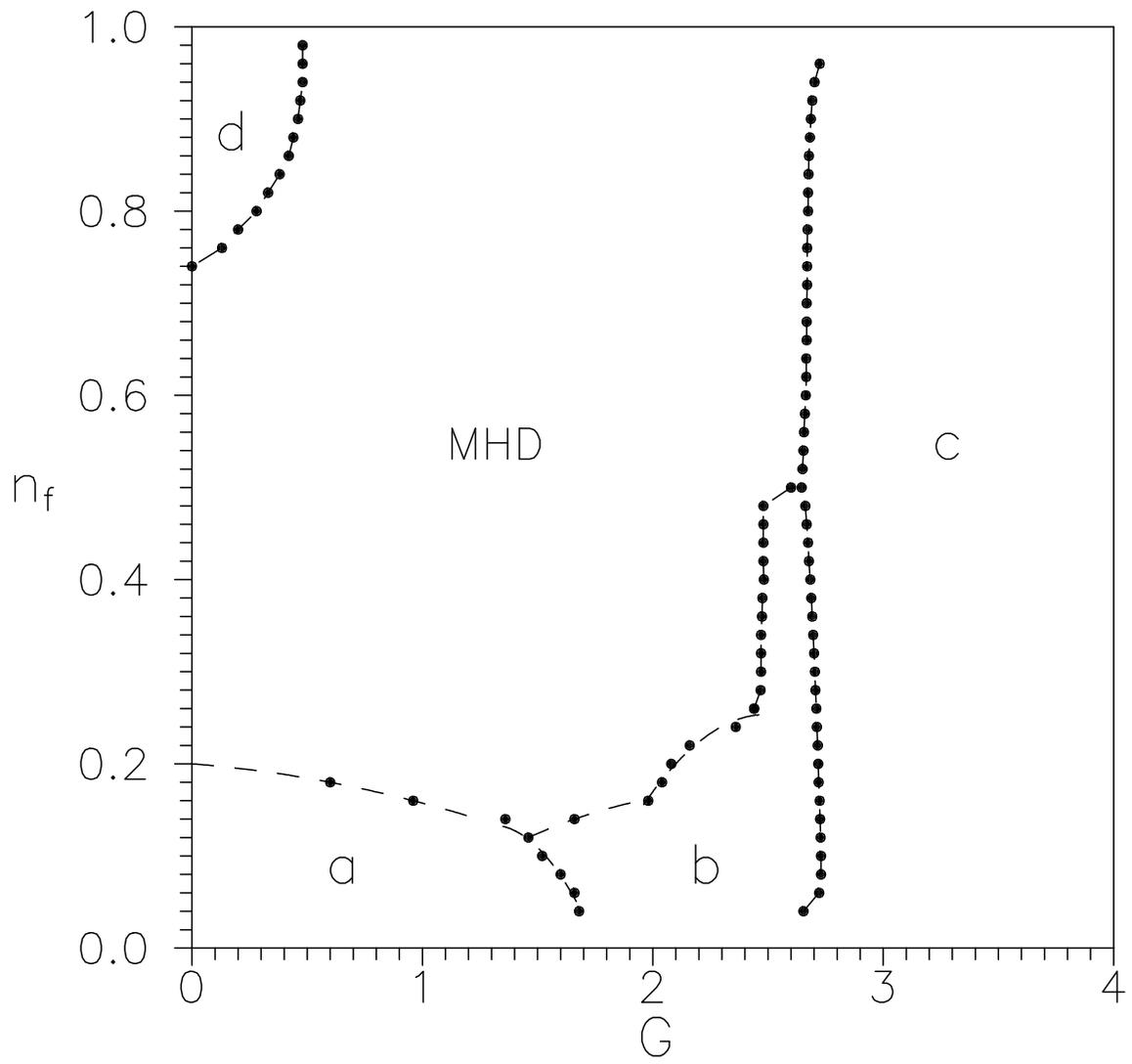
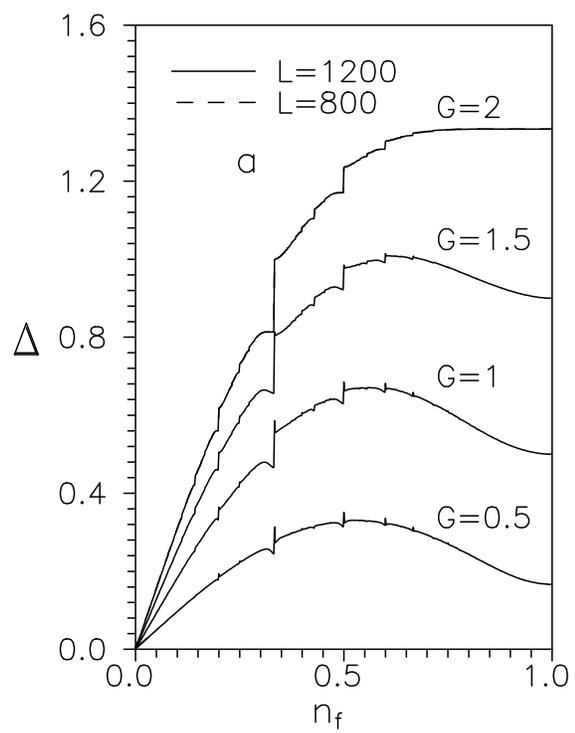
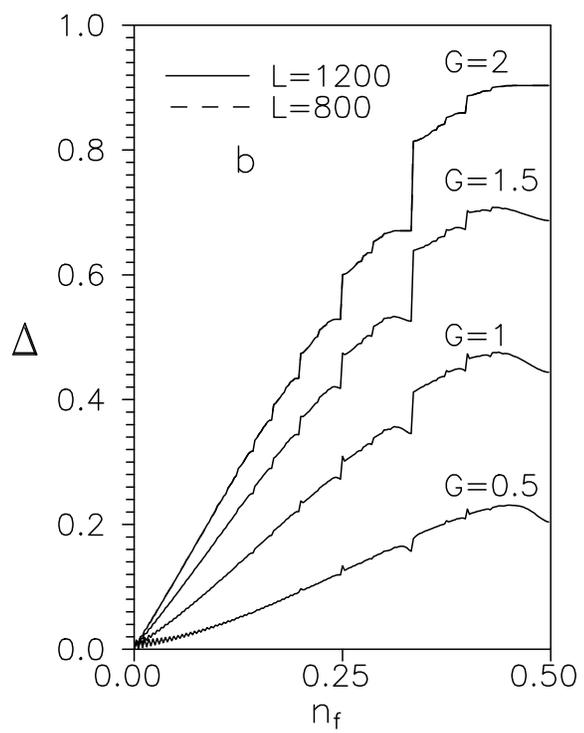
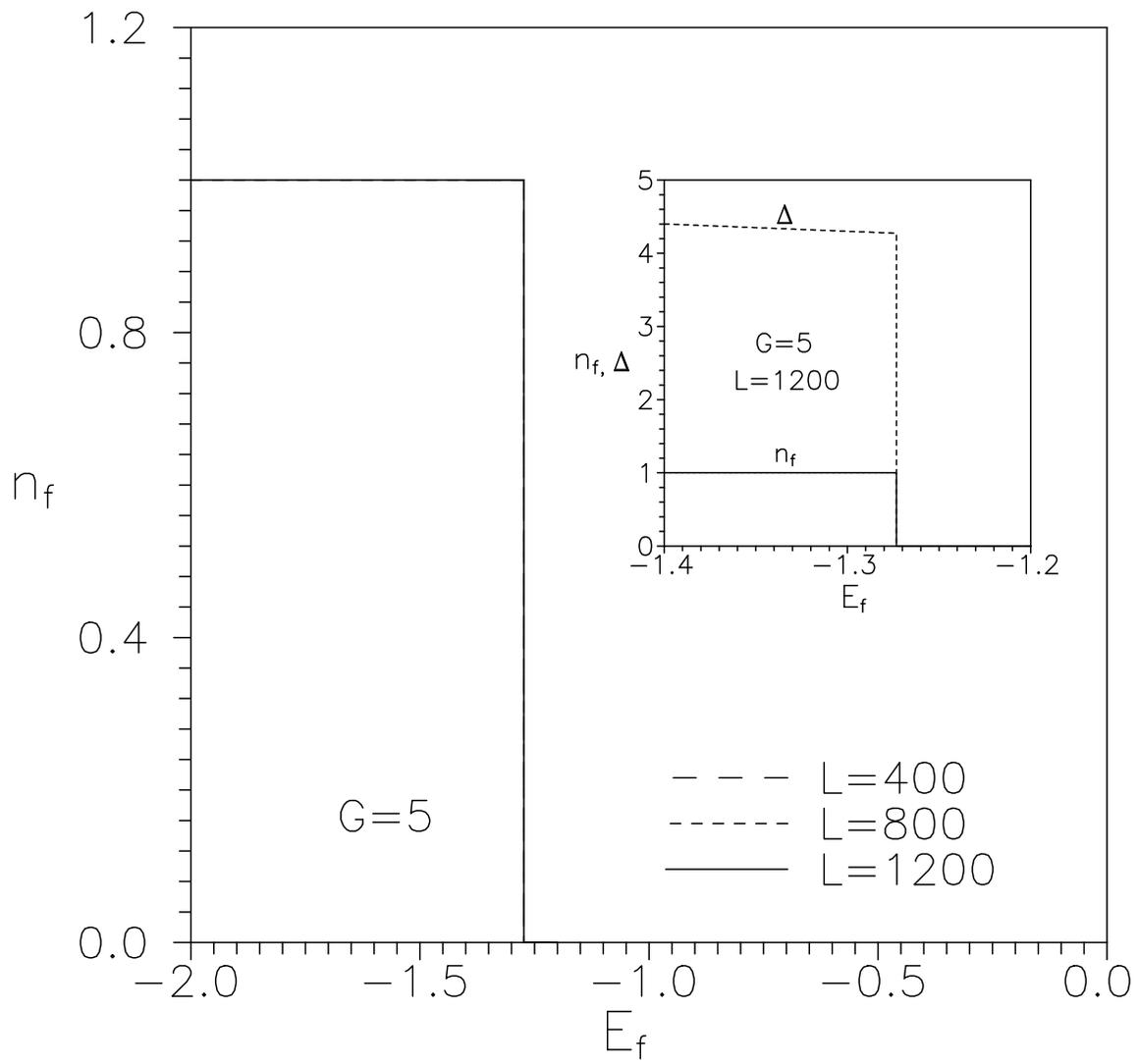
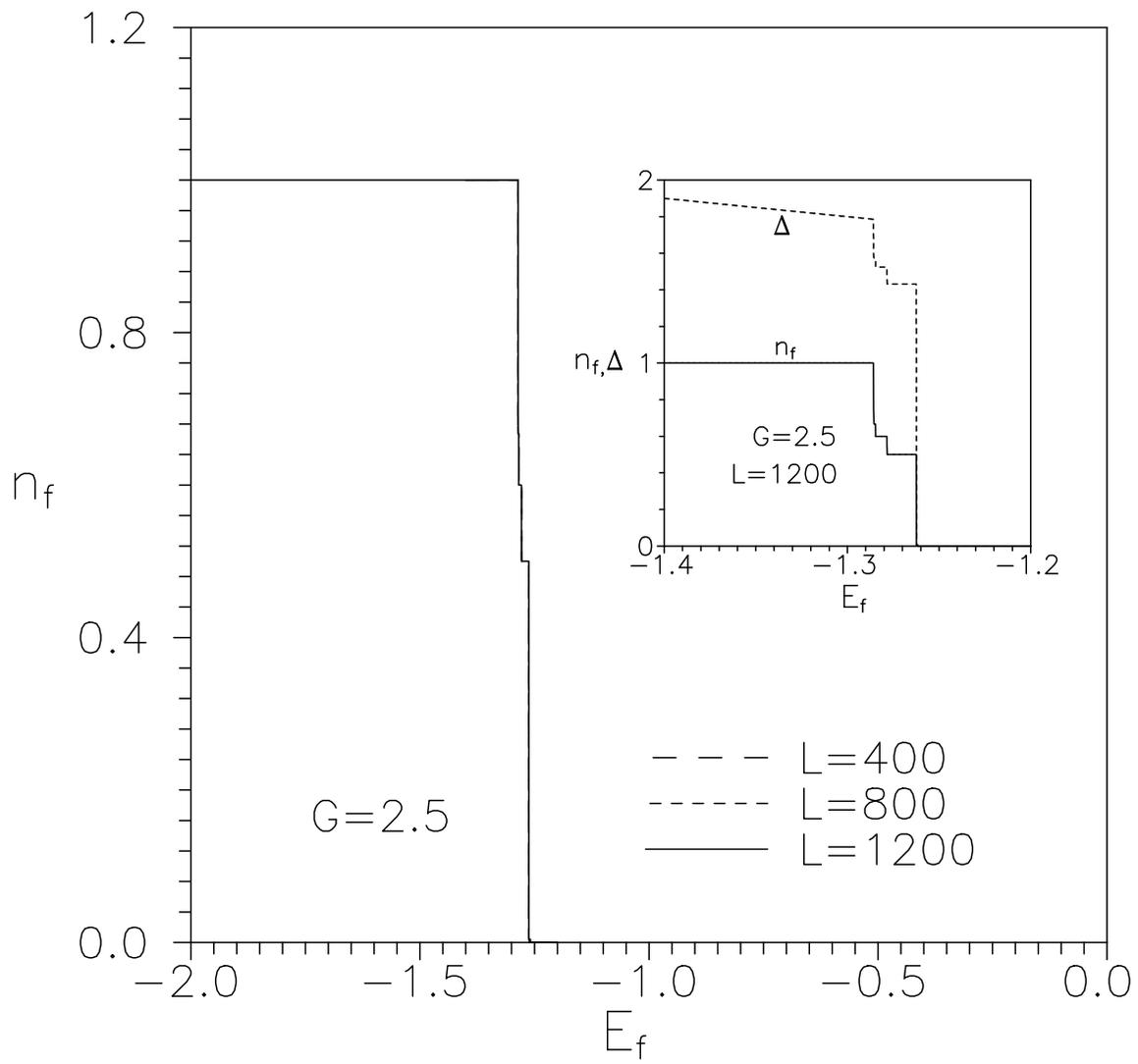
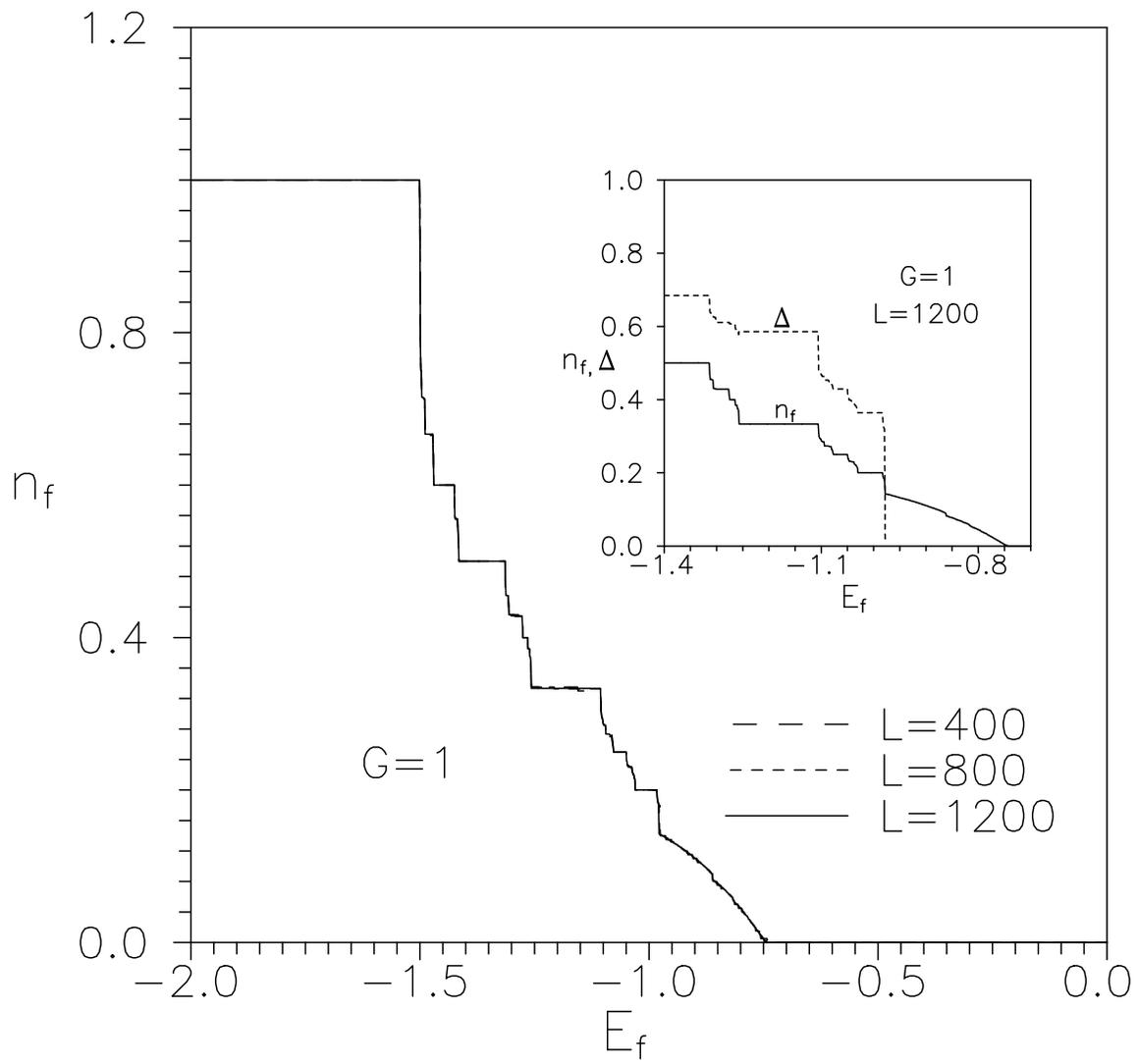
\end{document}